\documentclass[%
 reprint,
superscriptaddress,
 amsmath,amssymb,
 aps,
]{revtex4-2}

\usepackage{graphicx}
\usepackage{dcolumn}
\usepackage{bm}

\usepackage{aas_macros}
\usepackage{color}
\usepackage{amsmath}
\usepackage{natbib}

\usepackage{setspace}

\newcommand{\average}[1]{\ensuremath{\langle#1\rangle}}

\usepackage{hyperref}
\hypersetup{
    colorlinks=true,
    linkcolor=blue,
    citecolor=blue,        
    filecolor=magenta,      
    urlcolor=blue,
}

\begin{document}

\preprint{APS/123-QED}

\title{
Spin-flavor precession of Dirac neutrinos in dense matter and its potential in core-collapse supernovae
}


\author{Hirokazu Sasaki}
\email{hsasaki@lanl.gov}
\affiliation{Los Alamos National Laboratory, Los Alamos, New Mexico 87545, USA}

\author{Tomoya Takiwaki}
\affiliation{%
National Astronomical Observatory of Japan, \\
2-21-1 Osawa, Mitaka, Tokyo 181-8588, Japan}

\author{A. Baha Balantekin}
\affiliation{%
Department of Physics, University of Wisconsin, Madison, Wisconsin 53706, USA}




\date{\today}

\begin{abstract}

 We calculate the spin-flavor precession (SFP) of Dirac neutrinos induced by strong magnetic fields and finite neutrino magnetic moments in dense matter. As found in the case of Majorana neutrinos, the SFP of Dirac neutrinos is enhanced by the large magnetic field potential and suppressed by large matter potentials composed of the baryon density and the electron fraction. The SFP is possible irrespective of the large baryon density when the electron fraction is close to 1/3. The diagonal neutrino magnetic moments that are prohibited for Majorana neutrinos enable the spin precession of Dirac neutrinos without any flavor mixing. With supernova hydrodynamics simulation data, we discuss the possibility of the SFP of both Dirac and Majorana neutrinos in core-collapse supernovae. The SFP of Dirac neutrinos occurs at a radius where the electron fraction is 1/3. The required magnetic field of the proto-neutron star for the SFP is a few $10^{14}\,$G at any explosion time. For the Majorana neutrinos, the required magnetic field fluctuates from $10^{13}\,$G to $10^{15}\,$G. Such a fluctuation of the magnetic field is more sensitive to the numerical scheme of the neutrino transport in the supernova simulation. 


\end{abstract}

\maketitle


\section{Introduction}
 
Dirac versus Majorana nature remains one of biggest questions of neutrino physics. During the last several decades numerous experimental and theoretical efforts were directed to answer this question  \cite{Balantekin:2018}. Neutrinoless double beta ($0\nu\beta\beta$) decay is a promising probe into Majorana neutrinos and lower limits for the decay half-life on various nuclei were continuously updated by many  experiments \cite{0vbbreview2019}. Usual neutrino oscillation experiments can not distinguish between Dirac and Majorana neutrinos because the flavor conversions in vacuum and matter are associated with the differences of the squares of neutrino masses, not the indivudual masses. These experiments are not sensitive to the Majorana phases either \cite{Giunti:2007ry}. On the other hand, the coupling of stellar magnetic field and finite neutrino magnetic moments potentially induces the spin-flavor precession (SFP) between left-handed and right-handed neutrinos in astrophysical sites \cite{Lim:1987tk,Akhmedov:1988uk,Balantekin:1990jg}. Such a SFP is completely different from usual flavor conversions conserving the neutrino chirality and its properties are sensitive to Dirac versus Majorana natures. 

Recent experiments (e.g., XENONnT \cite{XENON:2022ltv}, CONUS \cite{CONUS:2022qbb}, Dresden-II reactor and COHERENT \cite{Coloma:2022avw}, and XMASS-I \cite{XMASS:2020zke}) constrained the value of neutrino magnetic moment. Among them, the XENONnT experiment gives the most stringent upper limit, $\mu_{\nu}<6.4\times10^{-12}\mu_{B}$ ($90\%$ C.L.) \cite{XENON:2022ltv}, where $\mu_{B}$ is the Bohr magneton. In addition to the terrestrial experiments, the value of the neutrino magnetic moment can be constrained from the thermal evolution of astronomical phenomena \cite{Arceo-Diaz:2015pva,Capozzi:2020,Mori:2020niw,Vassh:2015yza,Grohs:2023xwa}. Currently, neutrino energy loss in globular clusters imposes on the most
stringent value, $\mu_{\nu}<1.2\times10^{-12}\mu_{B}$ \cite{Capozzi:2020}. 

A strong magnetic field larger than $10^{12}$\,G is possible in explosive astrophysical sites such as core-collapse supernovae and neutron star mergers (see \cite{Burrows07,Takiwaki09,Scheidegger10,Moesta15,Sawai16,Matsumoto2020,Mastumoto2022,Bugli21,Bugli23,Obergaulinger20,Obergaulinger21,Aloy21,Obergaulinger22}
and the references therein). Large numbers of neutrinos produced there potentially contain information of the magnetic field effects on neutrino oscillations. We note that the neutrino flux from astrophysical sites can be affected by even small intergalactic and interstellar magnetic fields (nG--$\mu$G) \cite{Alok:2022pdn,2022arXiv221211287K}.

Currently operational neutrino observatories can detect of the order of $10^{4}$ neutrino events from a core-collapse supernova in the Galaxy \cite{Horiuchi2018WhatDetection}. Such high statistical neutrino signals enable the investigation on neutrino oscillations inside the star  \cite{Wu:2014kaa,Sasaki:2019jny,Zaizen2020JCAP,Zaizen:2020xum}. 
Furthermore, the nucleosynthesis of heavy nuclei induced by neutrino absorption such as the $\nu p$ process \cite{Frohlich:2005ys,Sasaki:2017jry,Sasaki:2021ffa,Saski2023arXiv230702785S} and $\nu$ process \cite{Woosley:1989bd,Hayakawa:2018ekx,Ko:2022uqv} is sensitive to large neutrino fluxes near the proto-neutron star (PNS). Therefore, observable quantities such as solar abundances and elemental abundances of heavy nuclei leave clues to the supernova neutrino \cite{Sasaki:2021ffa,Saski2023arXiv230702785S}.

It is predicted that neutrino oscillations in core-collapse supernovae are affected by the coherent forward scatterings of neutrinos with background matter and neutrinos themselves \cite{Volpe2023arXiv230111814V}. Charged current interactions of neutrinos with background electrons induce the Mikheyev–Smirnov–Wolfenstein (MSW) matter effect \cite{Wolfenstein1978NeutrinoMatter,Mikheev:1986gs}. The neutrino-neutrino interactions cause a nonlinear potential in the neutrino Hamiltonian \cite{Pantaleone:1992eq,Pantaleone:1992xh,Sigl:1992fn,McKellar:1992ja,Yamada:2000za,Balantekin:2006tg,Pehlivan:2011hp,Vlasenko:2013fja,Volpe:2013jgr,Blaschke:2016xxt} and give various properties of neutrino many-body system that are vigorously investigated such as the neutrino fast flavor instability \cite{Richers2022arXiv220703561R} and many-body nature of collective neutrino oscillation beyond the mean-field \cite{Patwardhan2023arXiv230100342P,Balantekin2023arXiv230501150B}. 

The SFP is affected by these potentials in core-collapse supernovae. The resonant spin-flavor (RSF) conversion occurs resonantly like the MSW effect at the resonance density in core-collapse supernovae \cite{Lim:1987tk,Akhmedov:1992ea,Akhmedov:1993sh,Totani:1996wf,Nunokawa:1996gp,Ando:2002sk,Ando:2003pj,Ando:2003is,Akhmedov:2003fu,Ahriche:2003wt,Yoshida:2009ec,Taygun2022arXiv220806926B}. Near the PNS, the neutrino-neutrino interactions are no longer negligible and contribute to neutrino--antineutrino oscillations of Majorana neutrinos \cite{deGouvea:2012hg,deGouvea:2013zp,Abbar:2020ggq,Sasaki:2021bvu}. Majorana neutrinos can reach flavor equilibrium 
in the short scale determined by the strength of magnetic field potential \cite{Abbar:2020ggq,Sasaki:2021bvu}. Such an equilibration phenomenon is induced by the coupling of matter potentials with the magnetic field potential and sensitive to the values of both the baryon density and the electron fraction of the supernova material. The neutrino--antineutrino oscillations can occur even with high baryon density if the electron fraction is close to $0.5$ \cite{Sasaki:2021bvu}. In our previous study (hereafter, Ref.~\cite{Sasaki:2021bvu} is denoted as ST21), we only focused on the equilibration of Majorana neutrinos. However, the SFP should also be possible for Dirac neutrinos. 

In this paper, we study the SFP of Dirac neutrinos in dense matter following a similar framework for Majorana neutrinos in ST21\cite{Sasaki:2021bvu}. We reveal the mechanism of equilibration of Dirac neutrinos and derive a necessary condition of the SFP in dense matter. Then, we investigate the possibility of SFP of both Dirac and Majorana neutrinos in core-collapse supernovae and discuss the difference between both types of neutrinos.

\section{ 
Methods
}\label{sec:methods}

\begin{figure*}
\includegraphics[width=1\linewidth]{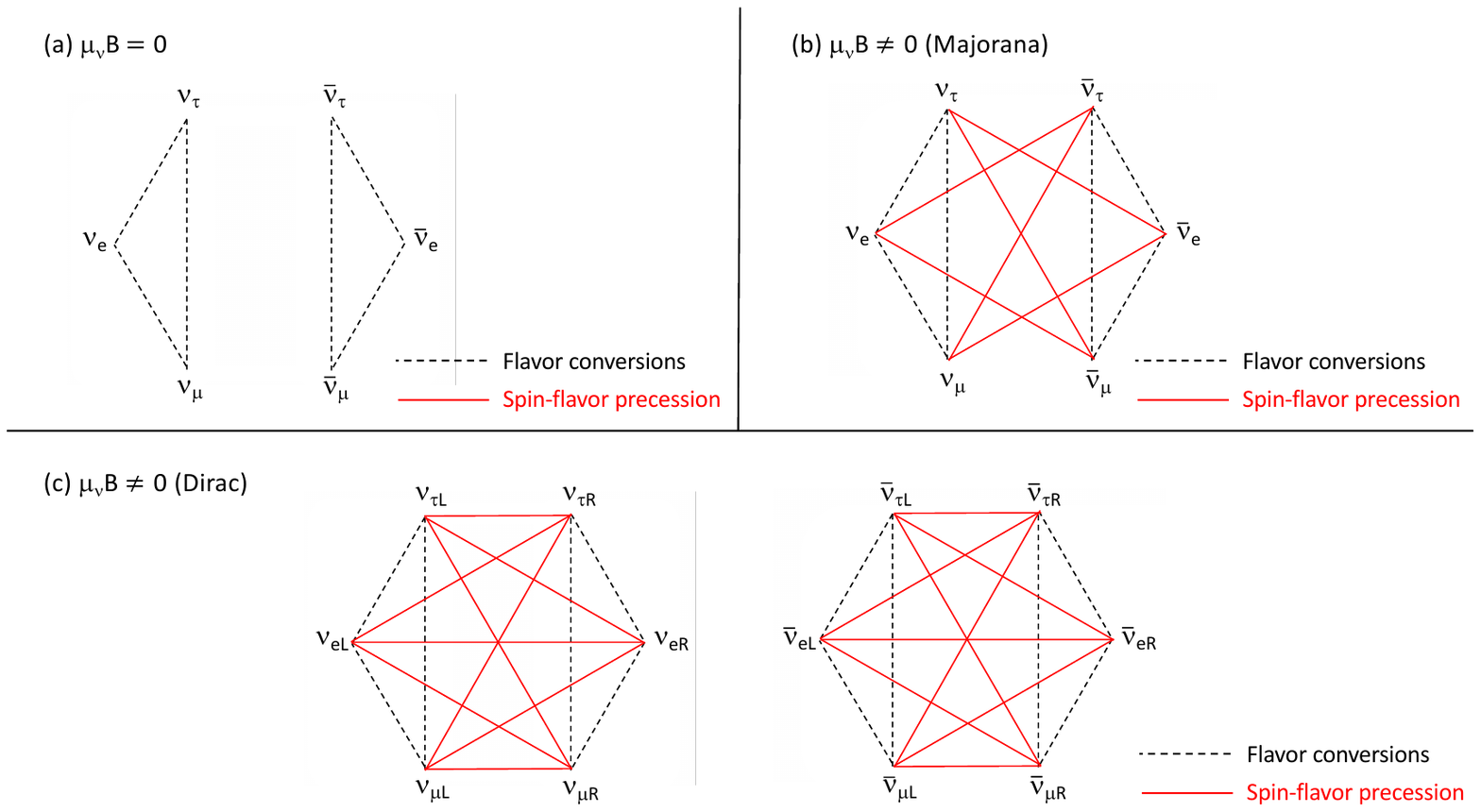}\\
\caption{
An overview of the magnetic field effect on neutrino oscillations. (a) the case without the magnetic field effect. The dashed black line shows ordinary flavor conversions in both neutrino and antineutrino sectors. (b) the case with the magnetic field effect on Majorana neutrinos. The solid red line shows the SFP induced by the finite neutrino magnetic moment and the magnetic field. (c) the magnetic field effect on Dirac neutrinos. The neutrinos and antineutrinos are decoupled with each other. $\nu_{\alpha L}$ and $\bar{\nu}_{\alpha R}(\alpha=e,\mu,\tau)$ correspond to active left-handed neutrinos ($\nu_{\alpha}$) and right-handed antineutrinos ($\bar{\nu}_{\alpha}$) in (a) and (b).}
\label{fig:nomajdir}
\end{figure*}

Figure~\ref{fig:nomajdir} gives an overview of standard neutrino flavor oscillations as well as the SFP.
In the absence of a magnetic field or a neutrino magnetic moment ($\mu_{\nu}B=0$), there is no spin precession among the left- and right-handed neutrinos and the only separate flavor conversions occur in the neutrino and the antineutrino sectors as illustrated in Fig.~\ref{fig:nomajdir}(a). When both the neutrino magnetic moment and magnetic field are finite ($\mu_{\nu}B\neq0$), then the SFP occurs together with usual flavor conversions. For the Majorana neutrinos in Fig.~\ref{fig:nomajdir}(b), the right-handed neutrino corresponds to the antineutrino and the SFP convert neutrinos into antineutrinos and vice versa. The spin precession without changing the flavor such as $\nu_{\alpha}\leftrightarrow\bar{\nu}_{\alpha}(\alpha=e,\mu,\tau)$ is prohibited due to the vanishing diagonal magnetic moments for Majorana neutrinos. With a strong magnetic field, Majorana neutrinos can reach an equilibrium of neutrino--antineutrino oscillations \cite{Abbar:2020ggq,Sasaki:2021bvu}. On the other hand, for the Dirac neutrinos, the neutrino and antineutrino sectors are decoupled as in Fig.~\ref{fig:nomajdir}(c), and the spin precession such as $\nu_{\alpha L}\leftrightarrow\nu_{\alpha R}$ is allowed due to the existence of diagonal magnetic moments. The magnetic field effect on neutrino oscillations is significantly different for Majorana and Dirac neutrinos although there is no difference in ordinary flavor conversions.

We study the magnetic field effect on Dirac neutrino oscillations as in Fig.~\ref{fig:nomajdir}(c) following our previous work on Majorana neutrino oscillations (ST21\cite{Sasaki:2021bvu}). We focus only on conversions of left-handed and right-handed neutrinos and ignore the effect of neutrino-neutrino interactions. We consider a single neutrino energy, $E=1\,$MeV with an emission angle $\theta\in[-\frac{\pi}{3},\frac{\pi}{3}]$ at a radius $r$ (This projection angle $\theta$ should not be confused with the neutrino mixing angles which carry indices, i.e. $\theta_{12}$.). 
The neutrino conversions are calculated with the Liouville von Neumann equation \cite{Abbar:2020ggq,Sasaki:2021bvu},
\begin{eqnarray}
\label{eq:liouville von neumann equation}
    \cos\theta\frac{\partial}{\partial r}D=-i[H,D],
\end{eqnarray}
where $D$ and $H$ are the neutrino density matrix and Hamiltonian for $\nu_{L}$ and $\nu_{R}$,
\begin{eqnarray}
  D&=&\left(
  \begin{array}{cc}
      \rho_{L\theta} &  X_{\theta}\\
      X_{\theta}^{\dagger} & \rho_{R\theta} 
  \end{array}\right),\label{eq:density matrix}
  \\
  H&=&\left(
  \begin{array}{cc}
      H_{L}&  V_{\mathrm{mag}}\\
       V_{\mathrm{mag}}^{\dagger}& H_{R}
  \end{array}\right), \label{eq:Hamiltonian}
\end{eqnarray}
\begin{equation}
\label{eq:Hamiltonian L}
    H_{L}=\Omega(E)+V_{\mathrm{mat}},
\end{equation}
where $\rho_{L\theta}$ and $\rho_{R\theta}$ are $3\times3$ density matrices of left- and right-handed neutrinos emitted with the angle  $\theta$, and $X_{\theta}$ is a correlation between $\nu_{L}$ and $\nu_{R}$. We use the vacuum term of left-handed neutrinos $\Omega(E)$ in Eq.~(8) of Ref.~\cite{Sasaki:2017jry} with neutrino mixing parameters, $\{\Delta m_{21}^{2},\Delta m_{32}^{2},\theta_{12},\theta_{13},\theta_{23}\}$, where we assume normal neutrino mass hierarchy ($\Delta m_{32}^{2}>0$) and no CP phase ($\delta_{\rm CP}=0$). The Hamiltonian of right-handed neutrinos is given by \cite{Lim:1987tk},
\begin{equation}
H_{R}
=\frac{\Delta m_{21}^{2}}{6E}\left(
\begin{array}{ccc}
     -2& 0 & 0\\
     0& 1 & 0 \\
     0& 0& 1
\end{array}
\right)+\frac{\Delta m_{32}^{2}}{6E}\left(
\begin{array}{ccc}
     -1& 0 & 0\\
     0& -1 & 0 \\
     0& 0& 2
\end{array}
\right),\
\end{equation}
where we ignore neutrino mixings and any interaction of the right-handed neutrinos with background particles. The matter potential in Eq.~(\ref{eq:Hamiltonian L}) is described by ST21\cite{Sasaki:2021bvu},
\begin{eqnarray}
    V_{\mathrm{mat}}&=&-\frac{\lambda_{n}}{2}I_{3\times3}
    +\lambda_{e}\left(
\begin{array}{ccc}
     1& 0 & 0\\
     0& 0 & 0 \\
     0& 0& 0
\end{array}
\right),\label{eq:lambdane}\\
\lambda_{e}&=&\sqrt{2}G_{F}\rho_{\mathrm{b}}N_{A}Y_{e},\\
\lambda_{n}&=&\sqrt{2}G_{F}\rho_{\mathrm{b}}N_{A}(1-Y_{e}),
\end{eqnarray}
where $G_{F}$ is the Fermi coupling constant, $\rho_{\mathrm{b}}$ is the baryon density, $N_{A}$ is the Avogadro number, $Y_{e}$ is the electron fraction, and $I_{3\times3}$ is the $3\times3$ identity matrix. $V_{\mathrm{mag}}$ in Eq.~(\ref{eq:Hamiltonian}) is the potential associated with the coupling of neutrino magnetic moments with the magnetic field,
\begin{eqnarray}
\label{eq:potential mag ori}
    V_{\mathrm{mag}}=
    B_{T}\left(
    \begin{array}{ccc}
     \mu_{ee}& \mu_{e\mu} & \mu_{e\tau}\\
     \mu_{\mu e}& \mu_{\mu\mu} & \mu_{\mu\tau} \\
     \mu_{\tau e}& \mu_{\tau\mu}& \mu_{\tau\tau}
\end{array}
\right),
\end{eqnarray}
where $B_{T}$ is a transverse magnetic field  perpendicular to the direction of neutrino emission and $\mu_{\alpha\beta}(\alpha,\beta=e,\mu,\tau)$ are neutrino magnetic moments. This magnetic field potential is in the non-diagonal component of Eq.~(\ref{eq:Hamiltonian}) inducing the mixing between left- and right-handed neutrinos. The diagonal neutrino magnetic moments $\mu_{\alpha\alpha}(\alpha=e,\mu,\tau)$ that are ignored in Majorana neutrinos need to be considered in Dirac neutrinos. Here we assume flavor-independent diagonal terms and anti-symmetric non-diagonal terms in Eq.~(\ref{eq:potential mag ori}),
\begin{eqnarray}
\label{eq:potential mag}
    V_{\mathrm{mag}}=
    \Omega_{\mathrm{d}}I_{3\times3}+\Omega_{\mathrm{nd}}\left(
    \begin{array}{ccc}
     0& 1& 1\\
     -1& 0 & 1 \\
     -1& -1& 0
\end{array}
\right),
\end{eqnarray}
where $\Omega_{\mathrm{d}}$ and $\Omega_{\mathrm{nd}}$ characterize the strengths of diagonal and non-diagonal neutrino magnetic moments.

\section{Results and Discussions}
\label{sec:check magnetic effect in supernovae}

\subsection{Non-diagonal magnetic field potential $\Omega_{\mathrm{nd}}$}

\label{sec:non-diagonal}

\begin{figure}[t]
\includegraphics[width=1\linewidth]{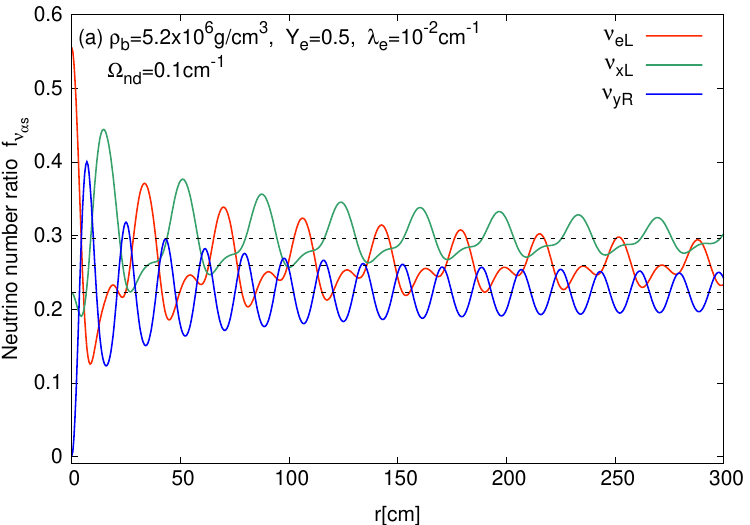}\\
\includegraphics[width=1\linewidth]{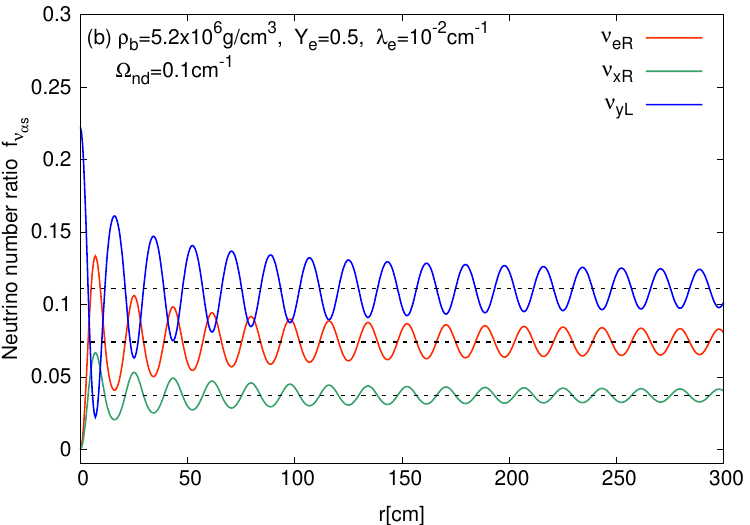}
\caption{
(a) The SFP of Dirac neutrinos with the non-diagonal term $\Omega_{\mathrm{nd}}=0.1\,\mathrm{cm}^{-1}$ in Eq.~(\ref{eq:potential mag}) for $\lambda_{e}=0.1\Omega_{\mathrm{nd}}$ and $Y_{e}=0.5$ in the $\nu_{eL}$--$\nu_{xL}$--$\nu_{yR}$ sector. The dashed lines show equilibrium values $(n_{\nu_{e}}+n_{\nu_{x}})/3n_{\nu}=7/27$, $(2n_{\nu_{e}}+3n_{\nu_{x}})/6n_{\nu}=8/27$, and $(2n_{\nu_{e}}+n_{\nu_{x}})/6n_{\nu}=2/9$. (b) The result of the $\nu_{eR}$--$\nu_{xR}$--$\nu_{yL}$ sector. The dashed lines correspond to the values of $n_{\nu_{x}}/3n_{\nu}=2/27$, $n_{\nu_{x}}/6n_{\nu}=1/27$, and $n_{\nu_{x}}/2n_{\nu}=1/9$.}
\label{fig:nondiag small matter}
\end{figure}

\begin{figure}[t]
\includegraphics[width=1\linewidth]{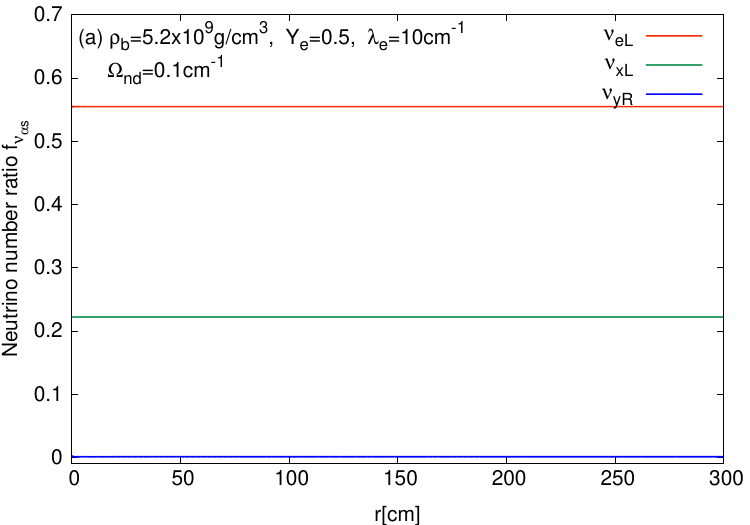}\\
\includegraphics[width=1\linewidth]{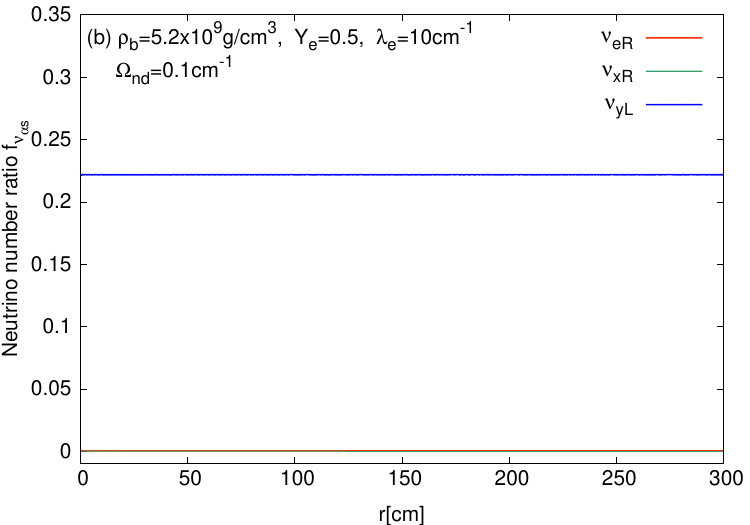}
\caption{
The SFP of Dirac neutrinos with the non-diagonal term $\Omega_{\mathrm{nd}}=0.1\,\mathrm{cm}^{-1}$ for $\lambda_{e}=10^{2}\Omega_{\mathrm{nd}}$ and $Y_{e}=0.5$ both in (a) $\nu_{eL}$--$\nu_{xL}$--$\nu_{yR}$ and (b) $\nu_{eR}$--$\nu_{xR}$--$\nu_{yL}$ sector.}
\label{fig:nondiag large matter}
\end{figure}

\begin{figure}[t]
\includegraphics[width=1\linewidth]{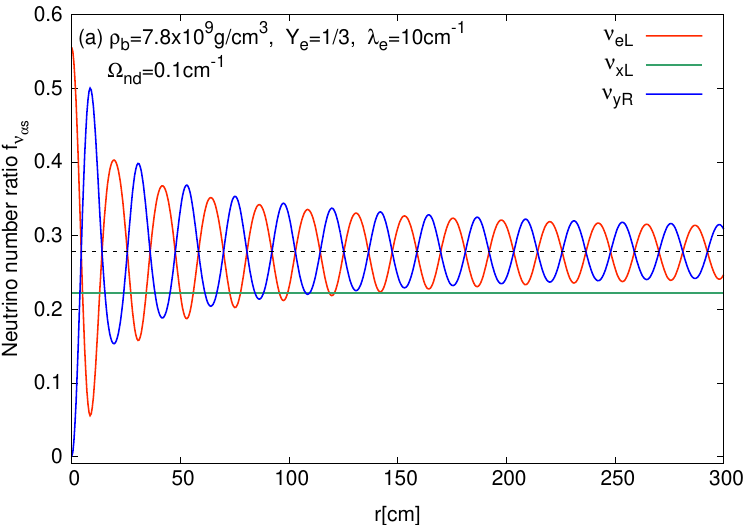}\\
\includegraphics[width=1\linewidth]{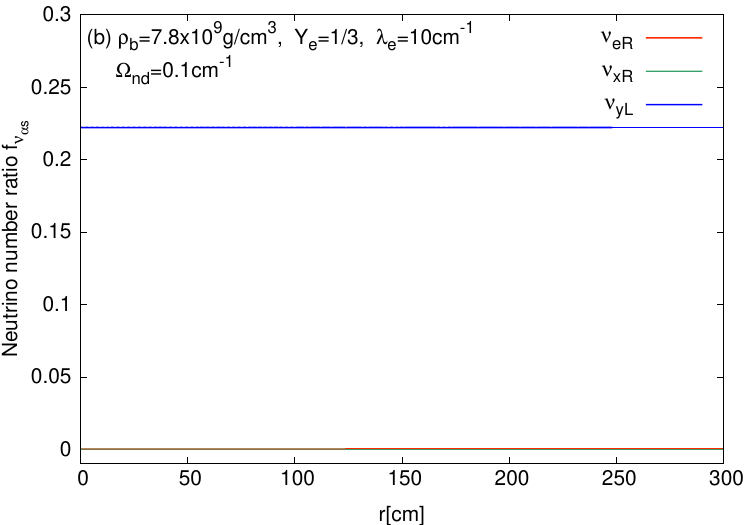}
\caption{
The SFP of Dirac neutrinos with the non-diagonal term $\Omega_{\mathrm{nd}}=0.1\,\mathrm{cm}^{-1}$ for $\lambda_{e}=10^{2}\Omega_{\mathrm{nd}}$ and $Y_{e}=1/3$ both in (a) $\nu_{eL}$--$\nu_{xL}$--$\nu_{yR}$ and (b) $\nu_{eR}$--$\nu_{xR}$--$\nu_{yL}$ sector. The dashed line in (a) shows $n_{\nu_{e}}/2n_{\nu}=5/18$.}
\label{fig:nondiag large matter ye=1/3}
\end{figure}

We calculate the magnetic field effect on Dirac neutrino oscillations by setting the initial condition of diagonal terms of left-handed neutrinos, $(\rho_{L\theta})_{\alpha\alpha}=n_{\nu_{\alpha}}/n_{\nu}\, (\alpha=e,\mu,\tau)$ where $n_{\nu}=\sum_{\alpha=e,\mu,\tau}n_{\nu_{\alpha}}$ and $n_{\nu_{x}}/n_{\nu_{e}}=0.4\, (x=\mu,\tau)$. All of the other components in Eq.~(\ref{eq:density matrix}) are set to zero at $r=0$. In Sec.~\ref{sec:non-diagonal}, we fix the strength of the non-diagonal magnetic field potential $\Omega_{\mathrm{nd}}=(\mu_{\nu}/10^{-12}\,\mu_{B})(B_{T}/3.4\times10^{14}\,\mathrm{G})=0.1\,\mathrm{cm}^{-1}$ and ignore the diagonal term ($\Omega_{\mathrm{d}}=0\,\mathrm{cm}^{-1}$). In order to study the sensitivity of matter potential in Eq.~(\ref{eq:lambdane}), we change the values of electron fraction $Y_{e}$ and the baryon density given by $\rho_{\mathrm{b}}=5.2\times10^{6}\,\mathrm{g}/\mathrm{cm}^{3}(Y_{e}/0.5)^{-1}(\lambda_{e}/10^{-2}\,\mathrm{cm}^{-1})$.

\paragraph{Low density case}
Figure~\ref{fig:nondiag small matter} shows the result of neutrino number ratios in the fixed values of $\rho_{\mathrm{b}}=5.2\times10^{6\,}\mathrm{g}/\mathrm{cm}^{3}$ and $Y_{e}=0.5$, which corresponds to $\lambda_{e}=0.1\Omega_{\mathrm{nd}}$. Such ratios of Dirac neutrinos $\nu_{\alpha s}(\alpha=e,x,y, s=L,R)$ are obtained by averaging the diagonal components of neutrino density matrices,
\begin{eqnarray}
    f_{\nu_{\alpha s}}
    &=&\frac{3}{2\pi}\int^{\frac{\pi}{3}}_{\frac{\pi}{3}}\mathrm{d}\theta\ (\rho_{s\theta})_{\alpha\alpha} ,
\end{eqnarray}
where $\nu_{xs}$ and $\nu_{ys}$ are eigenstates in a rotated frame described by a linear combination of flavors $\mu$ and $\tau$ \cite{Dasgupta:2007ws}. The magnetic field potential is a dominant term in neutrino Hamiltonian and SFPs in both $\nu_{eL}-\nu_{xL}-\nu_{yR}$ and $\nu_{eR}-\nu_{xR}-\nu_{yL}$ sectors are prominent. The vacuum frequencies associated with $\Delta m_{21}^{2}$ and $\Delta m_{32}^{2}$ are negligible compared with $\Omega_{\mathrm{nd}}$. Therefore, the values of $f_{\nu_{eL}}+f_{\nu_{xL}}+f_{\nu_{yR}}$ and $f_{\nu_{eR}}+f_{\nu_{xR}}+f_{\nu_{yL}}$ are almost constant. The neutrino ratios are oscillating around the dashed lines and such equilibrium values are derived in Appendix~\ref{sec:case1}.

The SFP discussed in Fig.~\ref{fig:nondiag small matter}--\ref{fig:diag} is almost independent of the vacuum term and the neutrino mass hierarchy. We remark that the vacuum frequencies can contribute to the SFP and the hierarchy dependence becomes prominent when the magnetic field potential is comparable with the vacuum frequencies and other potentials in the Hamiltonian are negligible. We mention such a hierarchy difference in core-collapse supernovae in Sec.~\ref{sec:dirac vs majorana}.

\paragraph{High density case}
Figure~\ref{fig:nondiag large matter} shows the result in the case of $\lambda_{e}=10^{2}\Omega_{\mathrm{nd}}$ with $\rho_{\mathrm{b}}=5.2\times10^{9\,}\mathrm{g}/\mathrm{cm}^{3}$ and $Y_{e}=0.5$. The numerical setup for Fig.~\ref{fig:nondiag large matter} is the same as that of Fig.~\ref{fig:nondiag small matter} except for the value of $\rho_{b}$. All of the calculated neutrino number ratios are constant ($f_{\nu_{eL}}=5/9,f_{\nu_{xL}}=f_{\nu_{yL}}=2/9,f_{\nu_{eR}}=f_{\nu_{xR}}=f_{\nu_{yR}}=0$) in Fig.~\ref{fig:nondiag large matter} and any SFP and usual flavor conversions are negligible. Such suppression due to the large matter potential is also confirmed in Majorana neutrinos \cite{Sasaki:2021bvu}. In general, the magnetic field potential should be dominant among the potentials in the neutrino Hamiltonian for the significant SFP.

\paragraph{Case for $Y_e\sim 1/3$}
The matter potentials $\lambda_{e}$ and $\lambda_{n}$ depends on both the baryon density and the electron fraction. Therefore, the matter suppression in Fig.~\ref{fig:nondiag large matter} is sensitive to the value of the electron fraction and the SFP is possible in some specific value of the electron fraction even in the large baryon density. Figure~\ref{fig:nondiag large matter ye=1/3} shows the result in 
$\lambda_{e}=10^{2}\Omega_{\mathrm{nd}}$ with $\rho_{\mathrm{b}}=7.8\times10^{9}\,\mathrm{g}/\mathrm{cm}^{3}$ and $Y_{e}=1/3$. Evolutions of $\nu_{xL},\nu_{yL},\nu_{eR},$ and $\nu_{xR}$ are negligible by the large matter potential $\lambda_{n}$. On the other hand, the SFP between $\nu_{eL}$ and $\nu_{yR}$ can avoid matter suppression. Such a SFP occurs where $|\lambda_{e}-\lambda_{n}/2|\ll \Omega_{\mathrm{nd}}$ is satisfied. $Y_{e}=1/3$ satisfies this condition irrespective of the value of $\rho_{b}$. We remark that, for Majorana neutrinos, a similar SFP independent of the value of $\rho_{b}$ occurs around $Y_{e}\sim0.5$ where $|\lambda_{e}-\lambda_{n}| \ll \Omega_{\mathrm{nd}}$ is satisfied \cite{Sasaki:2021bvu}. The mechanism of such $\rho_{\mathrm{b}}$ independent SFP of Dirac neutrinos at $Y_{e}=1/3$ is derived in Appendix~\ref{sec:case3}. More detail on the difference between Dirac and Majorana neutrinos is discussed in Appendix~\ref{sec:Majorana neutrinos}.

Almost the same discussion is possible for the right-handed antineutrinos by exchanging the sign of matter potential $\lambda_{e(n)}\to-\lambda_{e(n)}$. Then, the equilibrium values of active neutrinos ($f_{\nu_{\alpha}}^{\mathrm{eq}},f_{\bar{\nu}_{\alpha}}^{\mathrm{eq}}(\alpha=e,x,y)$) as in the dashed lines of Figs.~\ref{fig:nondiag small matter} and \ref{fig:nondiag large matter ye=1/3} are connected with initial neutrino ratios ($f_{\nu_{\alpha}}^{i},f_{\bar{\nu}_{\alpha}}^{i}$) through a transition matrix $U_{\mathrm{mag}}^{D}$,
\begin{equation}
    \left(
    \begin{array}{c}
     F^{\mathrm{eq}}_{\nu}\\
     F^{\mathrm{eq}}_{\bar{\nu}}
\end{array}
\right)=
\left(
    \begin{array}{cc}
U_{\mathrm{mag}}^{D}& 0\\
0& U_{\mathrm{mag}}^{D}
\end{array}
\right)
\left(
    \begin{array}{c}
     F^{\mathrm{i}}_{\nu}\\
     F^{\mathrm{i}}_{\bar{\nu}}
\end{array}
\right),\label{eq:transition matrix}
\end{equation}
\begin{eqnarray}
     F^{i(\mathrm{eq})}_{\nu}&=&\left(
    \begin{array}{c}
     f_{\nu_{e}}^{i(\mathrm{eq})}\\
     f_{\nu_{x}}^{i(\mathrm{eq})}\\
     f_{\nu_{y}}^{i(\mathrm{eq})}
\end{array}
\right),\label{eq:ieq flux n}\\
F^{i(\mathrm{eq})}_{\bar{\nu}}&=&\left(
    \begin{array}{c}
     f_{\bar{\nu}_{e}}^{i(\mathrm{eq})}\\
     f_{\bar{\nu}_{x}}^{i(\mathrm{eq})}\\
     f_{\bar{\nu}_{y}}^{i(\mathrm{eq})}
\end{array}
\right),\label{eq:ieq flux nb}
\end{eqnarray}
where $U_{\mathrm{mag}}^{D}$ is a $3\times3$ transition matrix. The values of the $U_{\mathrm{mag}}^{D}$ for three different extreme cases of neutrino potentials are shown in Table~\ref{tab:dirac nondiag}. Figs.~\ref{fig:nondiag small matter}--\ref{fig:nondiag large matter ye=1/3} correspond to the cases of (1)--(3) in the table. Our demonstration assumes that both the matter and the magnetic field potentials are sufficiently large compared with the vacuum Hamiltonian. For the case of (2) in Table~\ref{tab:dirac nondiag}, the SFP is suppressed and the $U_{\mathrm{mag}}^{D}$ is equivalent to the $3\times3$ identity matrix. For the case of (1) and (3), the summation of $f_{\nu_{\alpha}}^{\mathrm{eq}}(\alpha=e,x,y)$ is less than that of $f_{\nu_{\alpha}}^{i}$ due to the coupling with the right-handed neutrinos. Such a reduction of ratios of left-handed neutrinos is confirmed in antineutrinos too.

As shown in Eq.~(\ref{eq:transition matrix}), the neutrino and antineutrino sectors are decoupled from each other for Dirac neutrinos due to the decoupling as in Fig.~\ref{fig:nomajdir}(c). On the other hand, for Majorana neutrinos, the SFP occurs between neutrinos and antineutrinos as in Fig.~\ref{fig:nomajdir}(b). The equilibrium values of number ratios in three extreme cases are summarized in Eq.~(\ref{eq:transition matrix majorana}) and Table~\ref{tab:majorana nondiag} in Appendix~\ref{sec:Majorana neutrinos}. Since the total number of active neutrinos $(\sum_{\alpha=e,x,y}f_{\nu_{\alpha}}^{\mathrm{eq}}+f_{\bar{\nu}_{\alpha}}^{\mathrm{eq}})$ is conserved through the SFP in Majorana neutrinos, the reduced neutrino flux could be a prominent feature that distinguishes both Dirac and Majorana neutrinos.

\begin{table}
\begin{center}
\begin{tabular}{|c|c|}\hline
 Case&$U_{\mathrm{mag}}^{D}$ \\ \hline
(1) $|\lambda_{e}-\lambda_{n}/2|,\lambda_{n}\ll\Omega_{\mathrm{nd}}$&$
\left(
    \begin{array}{ccc}
     \frac{1}{3}& \frac{1}{3} & 0\\
     \frac{1}{3}& \frac{1}{2} & 0 \\
     0& 0& \frac{1}{2}
\end{array}
\right)$
\\ \hline
(2) $|\lambda_{e}-\lambda_{n}/2|,\lambda_{n}\gg\Omega_{\mathrm{nd}}$&$
\left(
    \begin{array}{ccc}
     1& 0 & 0\\
     0& 1 & 0 \\
     0& 0& 1
\end{array}
\right)$
\\ \hline
(3) $|\lambda_{e}-\lambda_{n}/2|\ll\Omega_{\mathrm{nd}},\lambda_{n}\gg\Omega_{\mathrm{nd}}$&$
\left(
    \begin{array}{ccc}
     \frac{1}{2}& 0 & 0\\
     0& 1 & 0 \\
     0& 0& 1
\end{array}
\right)$
\\ \hline
\end{tabular}
\end{center}
\small
\caption{The transition matrix $U_{\mathrm{mag}}^{D}$ in Eq.~(\ref{eq:transition matrix}) for Dirac neutrinos with the nondiagonal magnetic field potential $\Omega_{\mathrm{nd}}$ in Eq.~(\ref{eq:potential mag}) in three extreme cases of the matter and the magnetic field potentials.
Case (3) is satisfied if $Y_e\sim 1/3$.
}\label{tab:dirac nondiag}
\end{table}

\subsection{Diagonal magnetic field potential $\Omega_{\mathrm{d}}$}
\label{sec:dirac vs majorana}
\begin{figure}[t]
\includegraphics[width=0.8\linewidth]{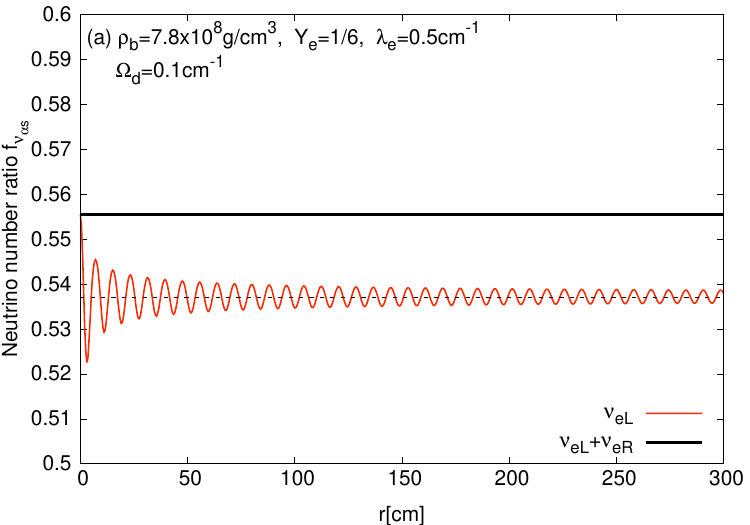}\\
\includegraphics[width=0.8\linewidth]{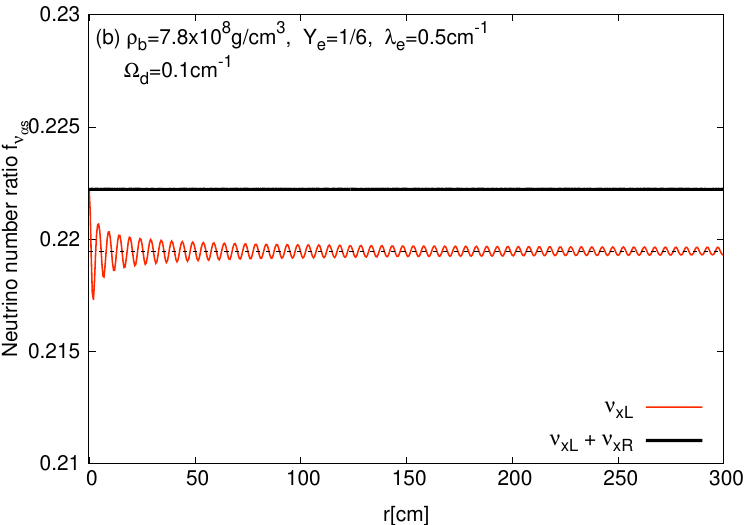}\\
\includegraphics[width=0.8\linewidth]{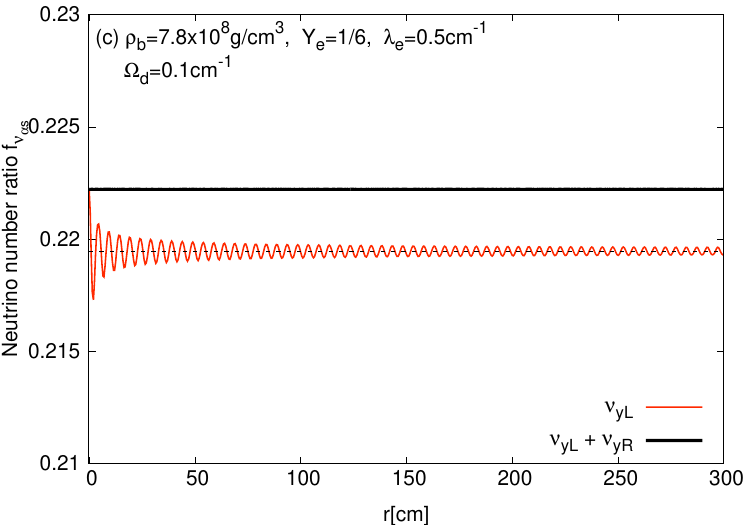}
\caption{
The SFP of Dirac neutrinos with the diagonal term $\Omega_{\mathrm{d}}=0.1\,\mathrm{cm}^{-1}$ in Eq.~(\ref{eq:potential mag}) for $\lambda_{e}=5\Omega_{\mathrm{d}}$ and $Y_{e}=1/6$ in (a) $\nu_{eL}$--$\nu_{eR}$, (b) $\nu_{xL}$--$\nu_{xR}$, and (c) $\nu_{yL}$--$\nu_{yR}$ sector. The dashed lines correpond to Eqs.~(\ref{eq:equilibrium dirac e}) and (\ref{eq:equilibrium dirac x}).}
\label{fig:diag}
\end{figure}
We discuss the SFP of Dirac neutrinos induced by the diagonal neutrino magnetic moment. We employ the same initial condition for the neutrino density matrix as Sec.~\ref{sec:non-diagonal} and fix the strength of magnetic field potential as $\Omega_{\mathrm{d}}=0.1\,\mathrm{cm}^{-1}$ and $\Omega_{\mathrm{nd}}=0\,\mathrm{cm}^{-1}$ in Eq.~(\ref{eq:potential mag}). 

Figure~\ref{fig:diag} shows the result of the SFP induced by diagonal magnetic field potential $\Omega_{\mathrm{d}}$ with $\rho_{\mathrm{b}}=7.8\times10^{8\,}\mathrm{g}/\mathrm{cm}^{3}$ and $Y_{e}=1/6$ resulting in $\lambda_{e}=5\Omega_{\mathrm{d}}$. The red lines are evolutions of $f_{\nu_{\alpha L}}(\alpha=e,x,y)$ and the black lines are the summations of $f_{\nu_{\alpha L}}+f_{\nu_{\alpha R}}$ that are almost constant in the calculation. Such constant black lines suggest that the SFPs occur in the same flavors $\nu_{\alpha L}$--$\nu_{\alpha R}(\alpha=e,x,y)$ and are decoupled with each other. The dynamics of $\nu_{\alpha L}$--$\nu_{\alpha R}$ can be solved in the same way as usual two flavor conversions. The result of Fig.~\ref{fig:diag}(a) is consistent with the conversion $\nu_{eL}$--$\nu_{eR}$ in Ref.~\cite{Lim:1987tk}. The equilibrium values of the left-handed neutrinos (dashed lines) are given by 
\begin{eqnarray}    
    f_{\nu_{e}}^{\mathrm{eq}}&=&f_{\nu_{e}}^{i}\left\{
    1-\frac{2\Omega_{\mathrm{d}}^{2}}{(\lambda_{e}-\frac{\lambda_{n}}{2})^{2}+4\Omega_{\mathrm{d}}^{2}}
    \right\},\label{eq:equilibrium dirac e}\\
    f_{\nu_{x(y)}}^{\mathrm{eq}}&=&f_{\nu_{x(y)}}^{i}\left\{
    1-\frac{2\Omega_{\mathrm{d}}^{2}}{(\frac{\lambda_{n}}{2})^{2}+4\Omega_{\mathrm{d}}^{2}}
    \right\}\label{eq:equilibrium dirac x}.
\end{eqnarray}
Similar relations are derived in the antineutrino sector. These equations show that conversions become maximum when the magnetic field potential is dominant $|\lambda_{e}-\lambda_{n}/2|,\lambda_{n}\ll\Omega_{\mathrm{d}}$. Conversely, any SFP is negligible when the matter potentials are large $|\lambda_{e}-\lambda_{n}/2|,\lambda_{n}\gg\Omega_{\mathrm{d}}$. The matter suppression does not occur in $\nu_{eL}$--$\nu_{eR}$ even in the large baryon density when the electron fraction is close to $1/3$ because of $|\lambda_{e}-\lambda_{n}/2|\sim0$. Our demonstration indicates that, in both non-diagonal and diagonal cases, the strong magnetic field potentials larger than $|\lambda_{e}-\lambda_{n}/2|$ are necessary for the significant SFP of Dirac neutrinos.

\subsection{Dirac vs Majorana in core-collapse supernovae}

We investigate the possibility of the SFP of both Dirac and Majorana neutrinos in core-collapse supernovae with neutrino spectra and matter profiles obtained in a neutrino radiation-hydrodynamic simulation of $11.2\, M_{\odot}$ progenitor model used in the demonstration for the Majorana neutrinos in ST21\cite{Sasaki:2021bvu}. 

As discussed in previous sections, the strength of magnetic field potential $\Omega_{\mathrm{mag}}=\mu_{\nu}B_{T}$ should be larger than $|\lambda_{e}-\lambda_{n}/2|(|\lambda_{e}-\lambda_{n}|)$ for the significant SFP of the Dirac (Majorana) neutrinos. To study the possibility of the SFP in core-collapse supernovae, we need to compare the magnetic field potential with other potentials in the neutrino Hamiltonian. 

Near the proto-neutron star (PNS) in core-collapse supernovae ($r\sim10$ km), the neutrino-neutrino interaction could induce a non-negligible potential in the neutrino Hamiltonian. The strength of the neutrino-neutrino interactions at the radius from the center $r$ could be estimated as ST21\cite{Sasaki:2021bvu},
\begin{equation}
\label{eq:zeta}
\begin{split}
\zeta=\frac{\sqrt{2}G_{F}}{4\pi R_{\nu}^{2}}\left|
\frac{L_{\nu_{e}}}{\average{E_{\nu_{e}}}}-\frac{L_{\bar{\nu}_{e}}}{\average{E_{\bar{\nu}_{e}}}}
\right|
\left(
1-\sqrt{1-\left(
\frac{R_{\nu}}{r}
\right)^{2}}
\right)^{2},
\end{split}
\end{equation}
where $\average{E_{\nu_{i}}}$ and $L_{\nu_{i}}$ are the mean energy and luminosity of $\nu_{i} (\nu_{i}=\nu_{e},\bar{\nu}_{e})$ on the surface of the PNS ($r=R_{\nu}$). We define the value of $R_{\nu}$ at a radius where the baryon density corresponds to $\rho_{\mathrm{b}}=10^{11}\,\mathrm{g}/\mathrm{cm}^{3}$. 

In outer region of supernova matter ($r>10^{3}$ km), the potentials of both matter and neutrino-neutrino interactions decrease and the vacuum Hamiltonian $\Omega(E)$ in Eq.~(\ref{eq:Hamiltonian L}) becomes dominant. The strength of the vacuum Hamiltonian in our demonstration could be characterized by the atmospheric vacuum frequency $\omega=|\Delta m_{32}^{2}|/2E$ where $E=10$ MeV is a typical neutrino mean energy in core-collapse supernovae.

The magnetic field potential should be dominant among the potentials in the neutrino Hamiltonian for the significant SFP. Then, the necessary conditions for the SFPs of both Dirac and Majorana neutrinos are given by
\begin{eqnarray}
\label{eq:necessary condition}
    \Omega_{\mathrm{mag}}&\geq&\eta \\
    \eta&=&\begin{cases}
\mathrm{max}\{\lambda_{D},\zeta,\omega \} & (\mathrm{Dirac})\\
\mathrm{max}\{\lambda_{M},\zeta,\omega \} & (\mathrm{Majorana}),
\end{cases}
\label{eq:eta}
\end{eqnarray}
where $\lambda_{D}=|\lambda_{e}-\lambda_{n}/2|$ and $\lambda_{M}=|\lambda_{e}-\lambda_{n}|$. $\eta$ is the maximum values among the three strengths and the difference between Dirac and Majorana neutrinos appears at $\lambda_{D}$ and $\lambda_{M}$.

For the model of the magnetic field, we use a dipole magnetic field as employed in ST21\cite{Sasaki:2021bvu},
\begin{equation}
\label{eq:magnetic field radial profile}
B_{T}=B_{0}\left(
\frac{R_{\nu}}{r}
\right)^{3},
\end{equation}
where $B_{0}$ is the transverse  magnetic field on the surface of PNS ($r=R_{\nu}$). The strength of magnetic field potential $\Omega_{\mathrm{mag}}=\mu_{\nu}B_{T}$ is calculated with a fixed value of the neutrino magnetic moment, $\mu_{\nu}=10^{-12}\mu_{B}$, satisfying the current experimental upper limit \cite{XENON:2022ltv}.

\begin{figure}[t]
\includegraphics[width=0.8\linewidth]{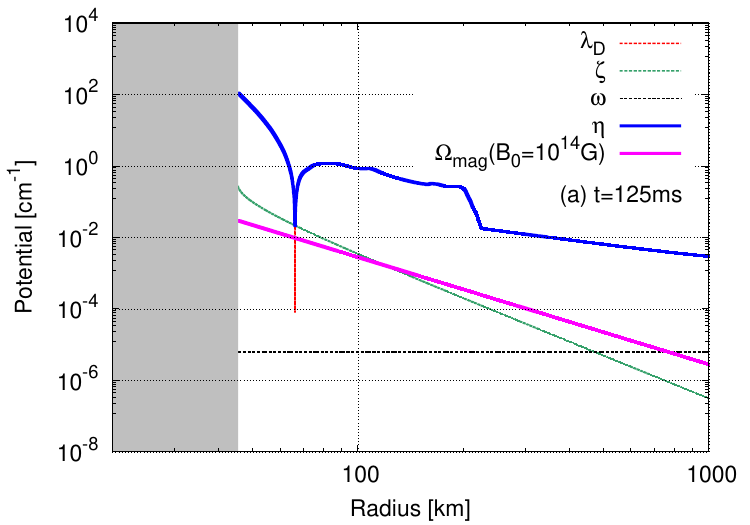}\\
\includegraphics[width=0.8\linewidth]{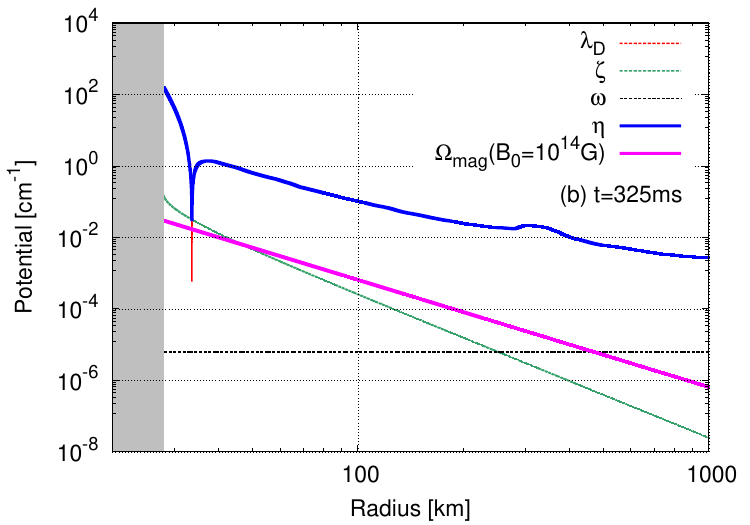}\\
\includegraphics[width=0.8\linewidth]{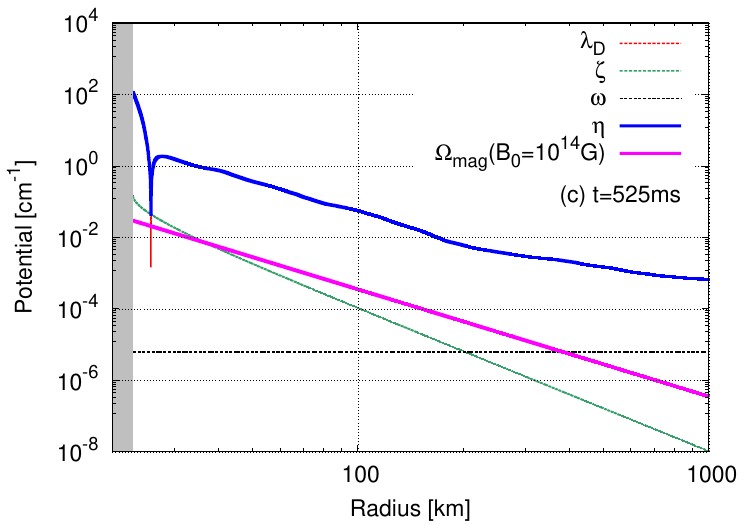}
\caption{
The potentials used for Eqs.~(\ref{eq:necessary condition}) and (\ref{eq:eta}) for Dirac neutrinos at different explosion time after postbounce ($t=125,\, 325,\, 525\,$ms). The shaded region shows the interior of the PNS.
}
\label{fig:criteria potential dirac}
\end{figure}

Figure~\ref{fig:criteria potential dirac} shows the strengths of the potentials and $\eta$ in Eq.~(\ref{eq:eta}) for Dirac neutrinos at $t=125,\, 325,\, 525$\,ms after postbounce. The shaded region shows the interior of the PNS. The radius of the boundary corresponds to the PNS radius $R_{\nu}$. The PNS radius becomes smaller as the explosion time has passed due to the shrink of the PNS. 
In all of the explosion time, the maximum strength $\eta$ (blue line) is given by $\lambda_{D}$ (red line) in $r>R_{\nu}$ except for a peak of $\lambda_{D}\sim0$ where $Y_{e}=1/3$ and $\eta=\zeta$ (green line) in Eq.~(\ref{eq:zeta}). Near the surface of PNS, $\lambda_{D}$ is large because of the dense and neutron-rich matter. On the other hand, the neutrino heating increases the value of electron fraction outside the PNS, which results in the point of $\lambda_{D}\sim0$ within $r<100\,$km. The baryon density decreases monotonously outside this point and the sudden decrease of $\lambda_{D}$ around 200 km in Fig.~\ref{fig:criteria potential dirac}(a) corresponds to the propagating shock front.

\begin{figure}[t]
\includegraphics[width=0.8\linewidth]{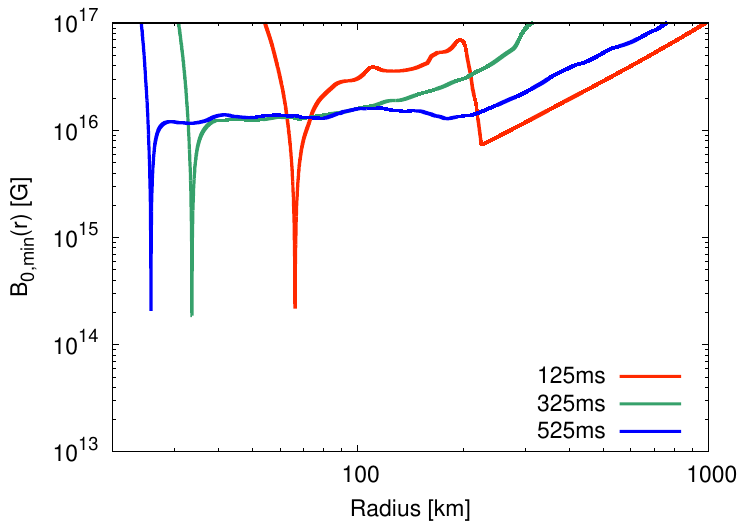}
\caption{
The radial profile of Eq.~(\ref{eq:B0min}) for Dirac neutrinos.
}
\label{fig:magnetic field radial profile dirac}
\end{figure}

The values of $\Omega_{\mathrm{mag}}$ in Fig.~\ref{fig:criteria potential dirac} (magenta lines) are obtained with the typical magnetic field of magnetars $B_{0}=10^{14}$ G in Eq.~(\ref{eq:magnetic field radial profile}). The magenta lines can move upwards with large $B_{0}$ keeping the same slope due to the same radial dependence of Eq.~(\ref{eq:magnetic field radial profile}). The necessary condition in Eq.~(\ref{eq:necessary condition}) is satisfied where the magenta lines are larger than the blue lines. The magenta lines in Fig.~\ref{fig:criteria potential dirac} are always smaller than the blue lines, which indicates that $B_{0}=10^{14}$ G is insufficient to satisfy Eq.~(\ref{eq:necessary condition}). For more quantitative discussion, we introduce a minimum magnetic field on the surface of PNS, 
\begin{equation}
\label{eq:B0min}
    B_{0,\mathrm{min}}(r)=\frac{r^{3}}{\mu_{\nu}R_{\nu}^{3}}\eta.
\end{equation}
Eq.~(\ref{eq:necessary condition}) is satisfied at the radius $r$ when $B_{0}>B_{0,\mathrm{min}}(r)$. Figure~\ref{fig:magnetic field radial profile dirac} shows the results of Eq.~(\ref{eq:B0min}) at different explosion times of Fig.~\ref{fig:criteria potential dirac}. The $B_{0,\mathrm{min}}(r)$ becomes minimum at the point of $\lambda_{D}\sim0$ at any explosion time. The minimum values are $2.24\times10^{14},\,1.86\times10^{14},\,2.23\times10^{14}\,$G at $t=125,\, 325,\, 525\,$ms after postbounce. The strong magnetic field $B_{0}>10^{16}\,$G is required for Eq.~(\ref{eq:necessary condition}) in regions other than such resonance point.

\begin{figure}[t]
\includegraphics[width=0.8\linewidth]{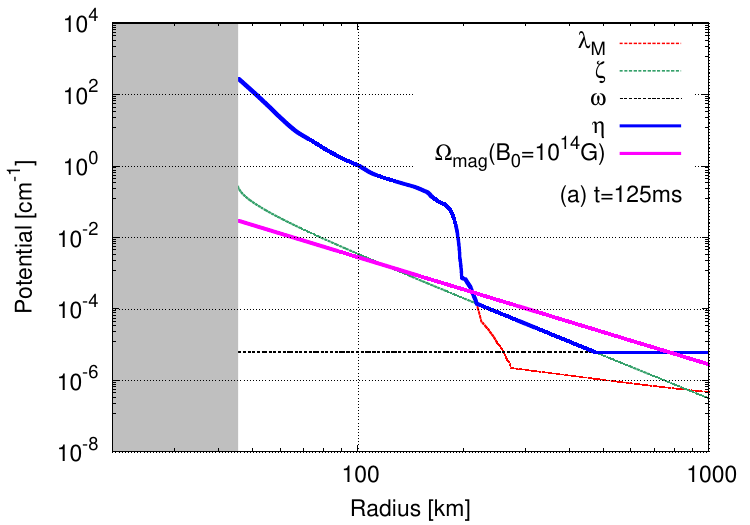}\\
\includegraphics[width=0.8\linewidth]{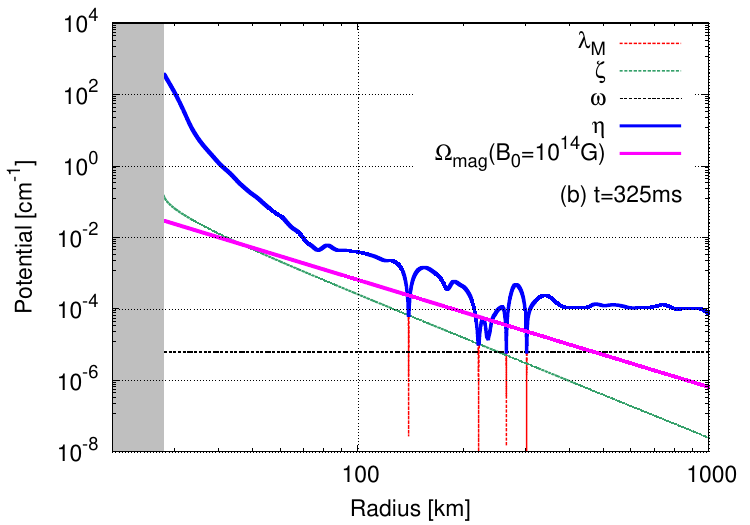}\\
\includegraphics[width=0.8\linewidth]{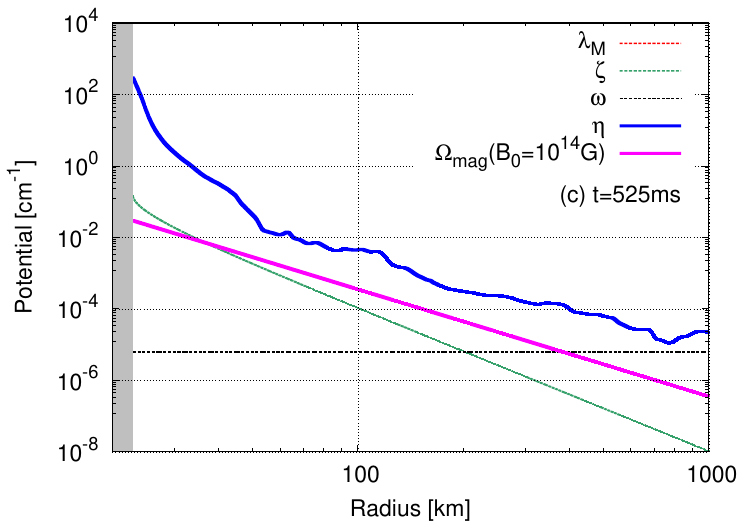}
\caption{
The potentials for the Majorana neutrinos in Eqs.~(\ref{eq:necessary condition}) and (\ref{eq:eta}) at $t=125,\, 325,\, 525\,$ms after postbounce as in Fig.~\ref{fig:criteria potential dirac}.
}
\label{fig:criteria potential moranaja}
\end{figure}

\begin{figure}[t]
\includegraphics[width=0.8\linewidth]{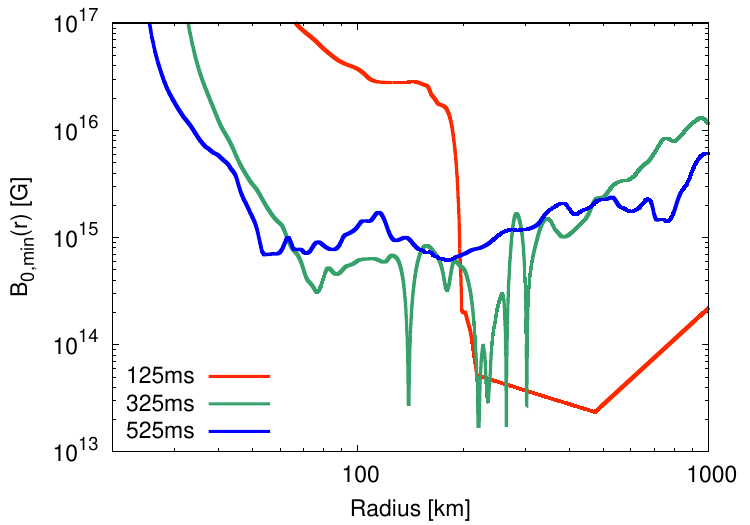}
\caption{
The radial profile of Eq.~(\ref{eq:B0min}) for Majorana neutrinos.
}
\label{fig:magnetic field radial profile majorana}
\end{figure}

Figure~\ref{fig:criteria potential moranaja} shows profiles of the potentials for Majorana neutrinos with different time snapshots as in Fig.~\ref{fig:criteria potential dirac}. In Fig.~\ref{fig:criteria potential moranaja} (a), at $t=125\,$ms, the maximum strength $\eta$ (blue line) is equal to $\lambda_{M}$ (red line) near the PNS radius and the $\lambda_{M}$ decreases significantly outside the shock front ($r>$200 km) due to the small baryon density and $Y_{e}\sim0.5$ in Si layer. Then, the $\eta$ is determined by $\zeta$ (green line) and $\omega$ (black line) in such outer region. The SFP around 800 km would depend on the neutrino mass hierarchy because the $\omega$ becomes maximum and comparable with $\Omega_{\mathrm{mag}}$. Such hierarchy dependence does not appear in Dirac neutrinos within 1000 km because of $\omega<\eta$ at any radius as shown in Fig.~\ref{fig:criteria potential dirac}(a). This does not happen even in $r>1000$ km because of $\omega\gg\Omega_{\mathrm{mag}}$. In Fig.~\ref{fig:criteria potential moranaja} (b), at $t=325\,$ms, there are several peaks of $\lambda_{M}\sim0$ (red line) where $Y_{e}=0.5$. Such peaks are favorable to satisfy Eq.~(\ref{eq:necessary condition}). The small $\lambda_{M}$ enables the crossing of $\Omega_{\mathrm{mag}}$ (magenta lines) and $\eta$ (blue lines) in Figs.~\ref{fig:criteria potential moranaja} (a) and \ref{fig:criteria potential moranaja} (b). Therefore, $B_{0}=10^{14}$ G is sufficient to satisfy Eq.~(\ref{eq:necessary condition}) for Majorana neutrinos. On the other hand, there are no peaks of $\lambda_{M}\sim0$ at $t=525\,$ms as in Fig.~\ref{fig:criteria potential moranaja} (c) because the electron fraction is not close to $0.5$. Then, there is no crossing of $\Omega_{\mathrm{mag}}$ (magenta line) and $\eta$ (blue line) in this explosion phase.

The result of Eq.~(\ref{eq:B0min}) for Majorana neutrinos is shown in Fig.~\ref{fig:magnetic field radial profile majorana}. As shown in the case of 125 ms (red line), the minimum value is given by $2.34\times10^{13}\,$G at $r=474\,$km in the Si layer outside the shock wave. After the shock propagation, $B_{0,\mathrm{min}}(r)$ in the shock-heated material becomes small at the point of $Y_{e}=0.5$ and the minimum value at 325 ms (green line) is given by $1.72\times10^{13}\,$G at $r=222\,$km. There is no point of $Y_{e}=0.5$ at 525 ms and $B_{0,\mathrm{min}}(r)$ is larger than $10^{14}$ G. 

\begin{figure}[t]
\includegraphics[width=1\linewidth]{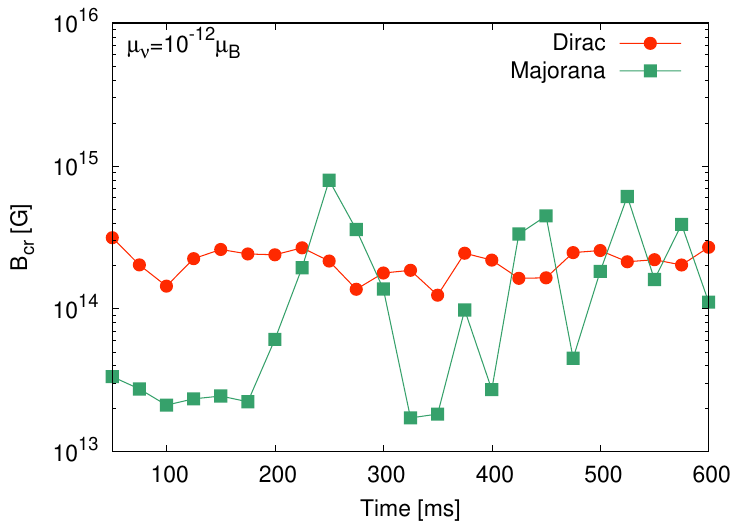}
\caption{
The critical magnetic field on the surface of the PNS in Eq.~(\ref{eq:critical magnetic field}) for both Dirac and Majorana neutrinos.
}
\label{fig:B0min}
\end{figure}

The minimum value of Eq.~(\ref{eq:B0min}) outside the PNS radius,
\begin{equation}
\label{eq:critical magnetic field}
     B_{\mathrm{cr}}=\min_{r\geq R_{\nu}}\left\{B_{0,\mathrm{min}}(r)
     \right\},
\end{equation}
could be regarded as a critical magnetic field on the surface of PNS for the SFP at the explosion time. 

Figure~\ref{fig:B0min} shows the result of Eq.~(\ref{eq:critical magnetic field}) at the different explosion times for both Dirac and Majorana neutrinos. The value for Dirac neutrinos is larger than $10^{14}$ G. As shown in Fig.~\ref{fig:magnetic field radial profile dirac}, the $B_{0,\mathrm{min}}(r)$ for Dirac neutrino becomes minimum at the radius of $Y_{e}\sim1/3$. Such a specific radius appears near the PNS radius inside the shock-heated material. We remark that the supernova matter profiles used in Figs.~\ref{fig:criteria potential dirac} and \ref{fig:criteria potential moranaja} are obtained by averaging the contribution from the various directions with the polar angle $\Theta\in[0,\pi]$ of the 2D hydrodynamic simulation. Hence the angular dependence on the profile of $Y_{e}$ is also averaged out in our analysis. In addition, our demonstration implicitly assumes the strength of the dipole magnetic field of the equator of the PNS ($\Theta=\pi/2$) and ignores the dependence of $\sin\Theta$ on Eq.~(\ref{eq:magnetic field radial profile}) \cite{Ando:2003pj}. Such a $\Theta$ dependence implies that the spin precession is less likely to occur in the polar direction ($\Theta\sim0$) due to the larger critical magnetic field.\\
\indent As shown in Fig.~\ref{fig:B0min}, the value of $B_{\mathrm{cr}}$ for Majorana neutrinos is less than $10^{14}$ G and smaller than those of Dirac neutrinos in $t<200$ ms. In this early explosion phase, as in Fig.~\ref{fig:criteria potential moranaja}(a), the value of $\lambda_{M}$ decreases significantly outside the shock front where $\rho_{b}$ is small and $Y_{e}$ is close to $0.5$ due to many $\alpha$ elements, which results in the small $B_{\mathrm{cr}}$. The SFP of Majorana neutrinos is more favorable than those of Dirac neutrinos in the early phase before the shock wave has reached the $\alpha$ elements layer.\\
\indent While propagating outwards, the shock wave heats the material of outer layers and changes the value of $Y_{e}$. In the later explosion phase ($t\geq200$ ms), $B_{\mathrm{cr}}$ for Majorana neutrinos fluctuates between $10^{13}$--$10^{15}$ G although the value for Dirac neutrinos is roughly constant. Such fluctuations originated from several peaks of $\lambda_{M}\sim0$ as in Fig.~\ref{fig:criteria potential moranaja}(b) and the absence of the peaks as in Fig.~\ref{fig:criteria potential moranaja}(c). The value of $Y_{e}$ is particularly uncertain around $0.5$ and dependent on the neutrino radiation transport scheme used in the supernova hydrodynamic simulation. More elaborate neutrino transport would enable a more detailed analysis of Majorana neutrinos in the later explosion phase.

In this study, we primarily focused on delineating the criteria for SFP, and omitted an elaboration of the anticipated neutrino event rate and spectrum.  To predict them, we need to investigate other aspects of neutrino oscillations, especially collective neutrino oscillations.
In particular, the implications of fast flavor conversions in supernova environments is actively debated (e.g., \cite{Duan2010,Tamborra2021,Capozzi2022,Nagakura2022}). A comprehensive study encompassing both SFP and fast flavor conversions is deferred to  future work.

\section{Conclusion} 
\label{sec:conclusion}

We calculate the SFP of Dirac neutrinos induced by the coupling of neutrino magnetic moments and magnetic fields in dense matter. This work is an extension of our previous work for neutrino--antineutrino oscillations of Majorana neutrinos \cite{Sasaki:2021bvu}. We demonstrate the SFP of Dirac neutrinos for both diagonal and non-diagonal neutrino magnetic moments. The SFP occurs significantly when the matter potentials are negligible compared with the magnetic field potential. The large baryon density tends to prevent the SFP. However, the SFP is possible in $\nu_{e}$ even with the large baryon density when the electron fraction is close to $1/3$. 

Finally, we verify the possibility of the SFP of both Dirac and Majorana neutrinos in core-collapse supernovae based on the necessary condition of the SFP derived from our demonstrations by using simulation data of $11.2\,M_{\odot}$ progenitor model with a fixed value of neutrino magnetic moment $\mu_{\nu}=10^{-12}\mu_{B}$. In the case of Dirac neutrinos, the required magnetic field on the surface of the PNS for the SFP is a few $10^{14}\,$G at any explosion time. The required magnetic field for Majorana neutrinos is a few $10^{13}\,$G in the early explosion phase before the shock reaches the Si layer. On the other hand, in the later explosion phase, the magnetic field fluctuates between $10^{13}$--$10^{15}\,$G, which is sensitive to the scheme of the neutrino transport used in hydrodynamic simulation.


\begin{acknowledgments}
This work was carried out under the auspices of the National Nuclear Security Administration of the U.S. Department of Energy at Los Alamos National Laboratory under Contract No. 89233218CNA000001. This study was supported in part by JSPS/MEXT KAKENHI Grant Numbers
Grant Number (
JP17H06364, 
JP21H01088, 
JP22H01223, 
JP23H01199  
and JP23K03400). 
This work is also supported by the NINS program for cross-disciplinary
study (Grant Numbers 01321802 and 01311904) on Turbulence, Transport,
and Heating Dynamics in Laboratory and Solar/Astrophysical Plasmas:
"SoLaBo-X”. This work was also supported in part by U.S. National Science Foundation Grants PHY-2020275 and PHY-2108339. 
Numerical computations were carried out on PC cluster and Cray XC50 at the Center for Computational Astrophysics,
National Astronomical Observatory of Japan.
This research was also supported by MEXT as “Program for Promoting 
researches on the Supercomputer Fugaku” (Toward a unified view of 
the universe: from large scale structures to planets) and JICFuS.

\end{acknowledgments}

\appendix

\section{Matter suppression with $\Omega_{\mathrm{nd}}$}
\label{sec:appendix analytical explanation}
Following a similar approach as done in ST21\cite{Sasaki:2021bvu}, we derive the equilibrium values (dashed lines) in Figs.~\ref{fig:nondiag small matter} and ~\ref{fig:nondiag large matter ye=1/3}. We solve Eq.~(\ref{eq:liouville von neumann equation}) with $\theta=0$ assuming the dense matter $H_{L}\sim V_{\mathrm{mat}}$, $H_{R}\sim0$, and the negligible diagonal neutrino magnetic moment $\Omega_{\mathrm{d}}=0$ in Eq.~(\ref{eq:potential mag}). The term $\theta$ in Eq.~(\ref{eq:density matrix}) is dropped hereafter. Then, the equation of motion in Eq.~(\ref{eq:liouville von neumann equation}) is decomposed by
\begin{eqnarray}
    \partial_{r}\rho_{L}&\sim&-i[V_{\mathrm{mat}},\rho_{L}]-i(V_{\mathrm{mag}}X^{\dagger}-XV_{\mathrm{mag}}^{\dagger}),\\
    \partial_{r}\rho_{R}&\sim&-i(V_{\mathrm{mag}}^{\dagger}X-X^{\dagger}V_{\mathrm{mag}}),\\
    \partial_{r}X&\sim&-i(V_{\mathrm{mat}}X+V_{\mathrm{mag}}\rho_{R}-\rho_{L}V_{\mathrm{mag}}),\label{eq:evolution X}
\end{eqnarray}
where $\partial_{r}=\partial/\partial_{r}$. The above equations of motion in $e-\mu-\tau$ basis can be described in $e-x-y$ basis through the transformation of matrices such as 
\begin{eqnarray}
V_{\mathrm{mat}}^{\prime}&=& R^{T}(\theta_{23})V_{\mathrm{mat}}R(\theta_{23})\nonumber\\
&=&V_{\mathrm{mat}},\label{eq:mat potential transform}\\
V_{\mathrm{mag}}^{\prime}&=& R^{T}(\theta_{23})V_{\mathrm{mag}}R(\theta_{23})\nonumber\\
&=&\Omega_{\mathrm{nd}}\left(
\begin{array}{c c c}
0&0&\sqrt{2}\\
0&0&1\\
-\sqrt{2}&-1&0
\end{array}\right),\label{eq:mag potential transform}\\
    R(\theta_{23})&=&\left(
\begin{array}{c c c}
1&0&0\\
0&\cos\theta_{23}&\sin\theta_{23}\\
0&-\sin\theta_{23}&\cos\theta_{23}
\end{array}\right),\label{eq:R23 transform}
\end{eqnarray}
where $\theta_{23}=\pi/4$. Here we comment that the matrix components of Eq.~(A4) in ST21\cite{Sasaki:2021bvu} should be corrected as Eq.~(\ref{eq:mag potential transform}) for a more precise discussion on Majorana neutrinos although the same equilibrium values of neutrino--antineutrino oscillations are derived with Eq.~(\ref{eq:mag potential transform}). Following the matrix transformation of Eqs.~(\ref{eq:mat potential transform})-(\ref{eq:R23 transform}), the evolutions of matrix components of the neutrino density matrices in $e-x-y$ basis are given by
\begin{eqnarray}
    \partial_{r}\rho_{Lee}&=&-2\sqrt{2}\Omega_{\mathrm{nd}}X_{ey,i},\label{eq:evolution rhoLee}\\
    \partial_{r}\rho_{Lxx}&=&-2\Omega_{\mathrm{nd}}X_{xy,i},\label{eq:evolution rhoLxx}\\
    \partial_{r}\rho_{Ryy}&=&2\sqrt{2}\Omega_{\mathrm{nd}}X_{ey,i}+2\Omega_{\mathrm{nd}}X_{xy,i},\\
    \partial_{r}\rho_{Lex}&=&-i\lambda_{e}\rho_{Lex}\nonumber\\
    &&-i\Omega_{\mathrm{nd}}(-X_{ey}+\sqrt{2}X_{xy}^{*}),\label{eq:evolution rhoLex}\\
    \nonumber\\
    \partial_{r}\rho_{Ree}&=&-2\sqrt{2}\Omega_{\mathrm{nd}}X_{ye,i},\\
    \partial_{r}\rho_{Rxx}&=&-2\Omega_{\mathrm{nd}}X_{yx,i},\\
    \partial_{r}\rho_{Lyy}&=&2\sqrt{2}\Omega_{\mathrm{nd}}X_{ye,i}+2\Omega_{\mathrm{nd}}X_{yx,i},\\
    \partial_{r}\rho_{Rex}&=&-i\Omega_{\mathrm{nd}}(X_{ye}^{*}-\sqrt{2}X_{yx}^{*}),\label{eq:evolution rhoRex}
\end{eqnarray}
where $X_{\alpha\beta,i}=\mathrm{Im}X_{\alpha\beta}(\alpha,\beta=e,x,y)$. From these equation of motions, conservation laws are obtained
\begin{eqnarray}
\label{eq:conservation laws}
\rho_{Lee}+\rho_{Lxx}+\rho_{Ryy}&=&\mathrm{const}.,\\
\rho_{Ree}+\rho_{Rxx}+\rho_{Lyy}&=&\mathrm{const}.,
\end{eqnarray}
which represents a decoupling of $\nu_{eL}-\nu_{xL}-\nu_{yR}$ and $\nu_{eR}-\nu_{xR}-\nu_{yL}$ sectors. We confirm that these conservation laws are numerically satisfied in Figs.~\ref{fig:nondiag small matter}--\ref{fig:nondiag large matter ye=1/3}. From Eq.~(\ref{eq:evolution X}), the evolution of $X$ is obtained and the evolutions of $X_{ey},X_{xy},X_{ye},X_{yx}$ are given by
\begin{eqnarray}
    \partial_{r}X_{ey}&=&-i\left(\lambda_{e}-\frac{\lambda_{n}}{2}\right)X_{ey}\nonumber\\
    &&-i\Omega_{\mathrm{nd}}\left\{
    \sqrt{2}(\rho_{Ryy}-\rho_{Lee})-\rho_{Lex}
    \right\},\label{eq:evolution Xey}\\
    \partial_{r}X_{xy}&=&-i\left(-\frac{\lambda_{n}}{2}\right)X_{xy}\nonumber\\
    &&-i\Omega_{\mathrm{nd}}\left(
    \rho_{Ryy}-\rho_{Lxx}-\sqrt{2}\rho_{Lex}^{*}
    \right),\label{eq:evolution Xxy}\\
    \nonumber\\
    \partial_{r}X_{ye}&=&-i\left(-\frac{\lambda_{n}}{2}\right)X_{ye}\nonumber\\
    &&-i\Omega_{\mathrm{nd}}\left\{
    \sqrt{2}(\rho_{Lyy}-\rho_{Ree})-\rho_{Rex}^{*}
    \right\},\\
    \partial_{r}X_{yx}&=&-i\left(-\frac{\lambda_{n}}{2}\right)X_{yx}\nonumber\\
    &&-i\Omega_{\mathrm{nd}}\left(
    \rho_{Lyy}-\rho_{Rxx}-\sqrt{2}\rho_{Rex}
    \right).
\end{eqnarray}
The above equations are needed to close the equations of Eqs.~(\ref{eq:evolution rhoLee})--(\ref{eq:evolution rhoRex}) and solved analytically in the extreme cases as in Table~\ref{tab:dirac nondiag}.

\subsection{Case of $|\lambda_{e}-\lambda_{n}/2|,\lambda_{n}\ll\Omega_{\mathrm{nd}}$}
\label{sec:case1}
In this case, the discussions on both $\nu_{eL}-\nu_{xL}-\nu_{yR}$ and $\nu_{eR}-\nu_{xR}-\nu_{yL}$ sectors are almost the same. Then, we only focus on the derivation in the $\nu_{eL}-\nu_{xL}-\nu_{yR}$ sector by using Eqs.~(\ref{eq:evolution rhoLee})--(\ref{eq:evolution rhoLex}) and (\ref{eq:evolution Xey})--(\ref{eq:evolution Xxy}). The second derivatives of $X_{ey,i}$ and $X_{xy,i}$ are described by
\begin{eqnarray}
    \partial_{r}^{2}\left(
\begin{array}{c c}
\sqrt{2}X_{ey,i}\\
X_{xy,i}
\end{array}\right)\sim-3\Omega_{\mathrm{nd}}^{2}\left(
\begin{array}{c c}
3&2\\
1&2
\end{array}\right)\left(
\begin{array}{c c}
\sqrt{2}X_{ey,i}\\
X_{xy,i}
\end{array}\right).
\end{eqnarray}
The above equation can be solved with the initial diagonal densities $\rho_{s\alpha\alpha}=\rho^{0}_{s\alpha\alpha}(\alpha=e,x,y,\, s=L,R)$ and the negligible correlation $X=0$ at $r=0$ as done in ST21\cite{Sasaki:2021bvu}. The diagonal components $\rho_{s\alpha\alpha}$ are composed of two modes, $\cos(\sqrt{3}\Omega_{\mathrm{nd}}r)$ and $\cos(2\sqrt{3}\Omega_{\mathrm{nd}}r)$ oscillating around the equilibrium values below
\begin{eqnarray}
    \rho^{\mathrm{eq}}_{Lee}&=&\frac{\rho^{0}_{Lee}+\rho^{0}_{Lxx}+\rho^{0}_{Ryy}}{3},\label{eq:eq rhoLee}\\
    \rho^{\mathrm{eq}}_{Lxx}&=&\frac{2\rho^{0}_{Lee}+3\rho^{0}_{Lxx}+\rho^{0}_{Ryy}}{6},\\
    \rho^{\mathrm{eq}}_{Ryy}&=&\frac{2\rho^{0}_{Lee}+\rho^{0}_{Lxx}+3\rho^{0}_{Ryy}}{6}.\label{eq:eq rhoRyy}
\end{eqnarray}
The dashed lines in Fig.~\ref{fig:nondiag small matter}(a) are reproduced with above equations by imposing $\rho^{0}_{Lee}=n_{\nu_{e}}/n_{\nu},\,\rho^{0}_{Lxx}=n_{\nu_{x}}/n_{\nu},\,$ and $\rho^{0}_{Ryy}=0$. The equilibrium values of the $\nu_{eR}-\nu_{xR}-\nu_{yL}$ sector in Fig.~\ref{fig:nondiag small matter}(b) are given by the replacements of $\rho_{L}\to\rho_{R}$ and $\rho_{R}\to\rho_{L}$ in Eqs.~(\ref{eq:eq rhoLee})--(\ref{eq:eq rhoRyy}).

\subsection{Case of $|\lambda_{e}-\lambda_{n}/2|,\lambda_{n}\gg\Omega_{\mathrm{nd}}$}
\label{sec:case2}
The second derivative of $X_{ey}$ and $X_{xy}$ are written as
\begin{eqnarray}
    \partial_{r}^{2}
X_{ey}&\sim&-\left(
\lambda_{e}-\frac{\lambda_{n}}{2}\right)^{2}
X_{ey},\\
\partial_{r}^{2}
X_{xy}&\sim&-\frac{\lambda_{n}^{2}}{4}
X_{xy}.
\end{eqnarray}
These equations are solved analytically. However, the negligible correlation $X=0$ at $r=0$ and Eqs.~(\ref{eq:evolution Xey})--(\ref{eq:evolution Xxy}) indicate $X_{ey}=X_{xy}=0$ irrespective of the radius, which results in no SFP as in Fig.~\ref{fig:nondiag large matter}. We can show negligible SFP in the $\nu_{eR}-\nu_{xR}-\nu_{yL}$ sector in the same way.

\subsection{Case of $|\lambda_{e}-\lambda_{n}/2|\ll\Omega_{\mathrm{nd}},\lambda_{n}\gg\Omega_{\mathrm{nd}}$}
\label{sec:case3}
In this case, there is no SFP in the $\nu_{eR}-\nu_{xR}-\nu_{yL}$ sector as in Sec.~\ref{sec:case2}. Such negligible SFP is consistent with numerical result in Fig.~\ref{fig:nondiag large matter ye=1/3}(b). In addition, $X_{xy}$ and $\rho_{Lex}$ are negligible due to large $\lambda_{n}$ and $\lambda_{e}$. From Eq.~(\ref{eq:evolution rhoLxx}), the value of $\rho_{Lxx}$ is constant and decoupled from the spin precesssion. Such a decoupling of $\nu_{xL}$ is confirmed numerically in Fig.~\ref{fig:nondiag large matter ye=1/3}(a). The second derivative of $X_{ey}$ is given by
\begin{eqnarray}
    \partial_{r}^{2}
X_{ey}&\sim&-8\Omega_{\mathrm{nd}}^{2}X_{ey}.
\end{eqnarray}
Following the same procedure as done in Sec.~\ref{sec:case1}, $\rho_{Lee}$ and $\rho_{Ryy}$ are described by a single mode of $\cos(2\sqrt{2}\Omega_{\mathrm{nd}}r)$ and the equilibrium value is written as
\begin{eqnarray}
    \rho^{\mathrm{eq}}_{Lee}=\rho^{\mathrm{eq}}_{Ryy}=\frac{\rho^{0}_{Lee}+\rho^{0}_{Ryy}}{2},
\end{eqnarray}
which reproduces the dashed line in Fig.~\ref{fig:nondiag large matter ye=1/3}(a) with $\rho^{0}_{Lee}=n_{\nu_{e}}/n_{\nu}$ and $\rho^{0}_{Ryy}=0$.

\section{Equilibrium values for Majorana neutrinos}
\label{sec:Majorana neutrinos}
A similar discussion can be made as in Sec.~\ref{sec:appendix analytical explanation} for Majorana neutrinos by replacing $H_{R}\to-V_{\mathrm{mat}}$ and $\rho_{R}\to\bar{\rho}_{R}$ and solving equations of motion below,
\begin{eqnarray}
    \partial_{r}\rho_{L}&\sim&-i[V_{\mathrm{mat}},\rho_{L}]-i(V_{\mathrm{mag}}X^{\dagger}-XV_{\mathrm{mag}}^{\dagger}),\\
    \partial_{r}\bar{\rho}_{R}&\sim&i[V_{\mathrm{mat}},\bar{\rho}_{R}]-i(V_{\mathrm{mag}}^{\dagger}X-X^{\dagger}V_{\mathrm{mag}}),\\
    \partial_{r}X&\sim&-i(V_{\mathrm{mat}}X+XV_{\mathrm{mat}}\nonumber\\
    &&+V_{\mathrm{mag}}\bar{\rho}_{R}-\rho_{L}V_{\mathrm{mag}}),\label{eq:evolution X majorana}
\end{eqnarray}
where $X$ is a correlation matrix between neutrinos and antineutrinos. The connection between equilibrium and initial fluxes in Eqs.~(\ref{eq:ieq flux n}) and (\ref{eq:ieq flux nb}) is given by
\begin{equation}
    \left(
    \begin{array}{c}
     F^{\mathrm{eq}}_{\nu}\\
     F^{\mathrm{eq}}_{\bar{\nu}}
\end{array}
\right)=U_{\mathrm{mag}}^{M}\left(
    \begin{array}{c}
     F^{i}_{\nu}\\
     F^{i}_{\bar{\nu}}
\end{array}
\right).\label{eq:transition matrix majorana}
\end{equation}
where $U_{\mathrm{mag}}^{M}$ is a $6\times6$ transition matrix due to the mixing of three flavors of neutrinos and those of antineutrinos as in Fig.~\ref{fig:nomajdir}(b). The values of three extreme cases for Majorana neutrinos are summarized in Table~\ref{tab:majorana nondiag}. Numerical results for Majorana neutrinos are shown in ST21\cite{Sasaki:2021bvu} and the equilibrium values of the three extreme cases are reproduced by $f_{\nu_{x}}^{i}=f_{\nu_{x}}^{i}=f_{\bar{\nu}_{x}}^{i}=f_{\bar{\nu}_{y}}^{i}$ in Eq.~(\ref{eq:transition matrix majorana}). We remark that $V_{\mathrm{mat}}X+XV_{\mathrm{mat}}$ in Eq.~(\ref{eq:evolution X majorana}) induces a frequency $\lambda_{e}-\lambda_{n}$ in Table~\ref{tab:majorana nondiag}. On the other hand, for Dirac neutrinos, a frequency $\lambda_{e}-\lambda_{n}/2$ in Table~\ref{tab:dirac nondiag} comes from $V_{\mathrm{mat}}X$ in Eq.~(\ref{eq:evolution X}). This difference for Majorana and Dirac neutrinos is the origin of the different necessary conditions in Eq.~(\ref{eq:eta}).

\begin{table}
\begin{center}
\begin{tabular}{|c|c|}\hline
 Case&$U_{\mathrm{mag}}^{M}$ \\ \hline
(1) $|\lambda_{e}-\lambda_{n}|,\lambda_{n}\ll\Omega_{\mathrm{nd}}$&$
\left(
    \begin{array}{cccccc}
     \frac{1}{3}& \frac{1}{3}& 0& 0& 0& \frac{1}{3}\\
     \frac{1}{3}& \frac{1}{2}& 0& 0& 0& \frac{1}{6}\\
     0& 0& \frac{1}{2}& \frac{1}{3}& \frac{1}{6}& 0\\
     0& 0& \frac{1}{3}& \frac{1}{3}& \frac{1}{3}& 0\\
     0& 0& \frac{1}{6}& \frac{1}{3}& \frac{1}{2}& 0\\
     \frac{1}{3}& \frac{1}{6}& 0& 0& 0& \frac{1}{2}\\
\end{array}
\right)$
\\ \hline
(2) $|\lambda_{e}-\lambda_{n}|,\lambda_{n}\gg\Omega_{\mathrm{nd}}$&$
\left(
    \begin{array}{cccccc}
     1& 0& 0& 0& 0& 0\\
     0& 1& 0& 0& 0& 0\\
     0& 0& 1& 0& 0& 0\\
     0& 0& 0& 1& 0& 0\\
     0& 0& 0& 0& 1& 0\\
     0& 0& 0& 0& 0& 1\\
\end{array}
\right)$
\\ \hline
(3) $|\lambda_{e}-\lambda_{n}|\ll\Omega_{\mathrm{nd}},\lambda_{n}\gg\Omega_{\mathrm{nd}}$&$
\left(
    \begin{array}{cccccc}
     \frac{1}{2}& 0& 0& 0& 0& \frac{1}{2}\\
     0& 1& 0& 0& 0& 0\\
     0& 0& \frac{1}{2}& \frac{1}{2}& 0& 0\\
     0& 0& \frac{1}{2}& \frac{1}{2}& 0& 0\\
     0& 0& 0& 0& 1& 0\\
     \frac{1}{2}& 0& 0& 0& 0& \frac{1}{2}\\
\end{array}
\right)$
\\ \hline
\end{tabular}
\end{center}
\small
\caption{The transition matrix $U_{\mathrm{mag}}^{M}$ in Eq.~(\ref{eq:transition matrix majorana}) for Majorana neutrinos in three extreme cases. Case (3) is satisfied if $Y_e\sim 0.5$.}
\label{tab:majorana nondiag}
\end{table}

\bibliography{ref}


\end{document}